\documentclass[]{spie}  

\usepackage{amsmath,amsfonts,amssymb}
\usepackage{graphicx,braket}
\usepackage[colorlinks=true, allcolors=blue]{hyperref}

\newcommand\D{\!\operatorname{d}\!}
\newcommand\rC{r_\text{\tiny C}}
\newcommand{\erf}{\operatorname{erf}}

\title{Collapse models and gravitational decoherence at test: How far can we push the limits of quantum mechanics?}

\author[a]{Matteo Carlesso}

\affil[a]{Centre for Quantum Materials and Technologies,
School of Mathematics and Physics, Queens University, Belfast BT7 1NN, United Kingdom}

\authorinfo{Further author information: (Send correspondence to M.C.)\\M.C.: E-mail: m.carlesso@qub.ac.uk}

\begin{document} 
\maketitle

\begin{abstract}
Collapse models  describe the breakdown of the quantum superposition principle when moving from microscopic to macroscopic scales. They are 
among the possible solutions to the quantum measurement problem and thus describe the emergence of classical mechanics from the quantum one. Testing collapse models is equivalent to test the limits of quantum mechanics. 
I will provide an overview  on how one can test collapse models, and which are the future theoretical and experimental challenges ahead.
\end{abstract}


\section{INTRODUCTION}
\label{sec:intro}  

The radical difference between classical and quantum mechanics is embedded in the quantum superposition principle, which allows a system to be in two or more different states at once. Such a building block of quantum theory is  valid and well tested in the  microscopic domain, but it has not being observed at macroscopic regimes. The breakdown of the quantum superposition principle is a major open question, and imposes limits to the validity of quantum theory.\\

One of the proposed solutions is provided by collapse models, which are phenomenological models describing a progressive loss of quantum coherence when the mass and complexity of the system increase \cite{bassi2003dynamical,bassi2013models}.  Such models suitably modify the Schr\"odinger dynamics so that the effects of the modifications are negligible for microscopic systems and are strong for macroscopic ones.
In such a way, they provide a smooth transition
from the micro-world, well described by quantum mechanics, to the macro-world,
where systems are never observed in superpositions. 
This explains the quantum-to-classical transition in a coherent way,
avoiding paradoxes like the famous Schr\"odinger’s cat. Thanks to the technological developments,  current experiments are now able to test  the
boundaries between the classical and quantum realms, thus providing  strong insights to collapse models and the limits of quantum mechanics\cite{arndt2014testing,carlessoNatPhys}.

\section{Collapse equation}

Collapse models add
stochastic and non-linear terms to the Schr\"odinger equation so that the
collapse of the quantum wavefunction is embedded in the dynamics. This
solves the quantum measurements problem, with no need of introducing a second evolution for  the wavepacket
reduction postulate.
The collapse models' modification of the Schr\"odinger equation reads as follows\cite{gisin1989stoch}:
\begin{equation}
\label{csl}
\D \ket{\psi_t}= \left[- \frac{i}{\hbar} \hat{H} \D t+ \int \D^3 {\bf x}\left(\hat{M}({\bf x})-\braket{ \hat{M}({\bf x}) }_t \right)\, 
\D W_t({\bf x})
-  \frac{1}{2}\int \D^3{\bf x}\D^3{\bf y}\,{\mathcal D}({\bf x}-{\bf y})\!\!\prod_{{\bf q}={\bf x},{\bf y}}\!\!
\left(\hat{M}({\bf q})-\braket{ \hat{M}({\bf q}) }_t \right)
\,\D t  \right] \ket{\psi_t},
\end{equation}
where $\hbar$ is the reduced Planck constant; $\hat H$ is the standard quantum Hamiltonian that leads to the standard Schr\"odinger equation.   The second and third terms describe the 
stochastic and non-linear modifications weighted by the mass density operator $\hat M({\bf x})$, which ensures a space localisation of  the wavefunction.
The Brownian noise $W_t({\bf x})$ with
spatial correlation equal to $D({\bf x}-{\bf y})$ and the non-linear term $\braket{\hat M({\bf x})}=\braket{\psi_t|\hat M({\bf x})|\psi_t}$
drive the collapse process.\\

The structure of the collapse equation ensures that it is norm-preserving although not being unitary\cite{gisin1989stoch}. Moreover, an amplification mechanism is automatically implemented: the collapse terms are proportional to the mass density operator $\hat M({\bf x})$ which makes the collapse rate of an object to scale roughly with its size. As a consequence, for microscopic systems, one has extremely small values for the collapse rate, thus re-establishing the standard quantum mechanical dynamics. On the other hand, macroscopic systems, through such an amplification  mechanism, are strongly affected and remain well localised in space.
Moreover, when a microscopic system is \textit{measured} by a macroscopic measurement device -- therefore an interaction between the two is assumed -- the collapse dynamics ensures that the outcomes at the end of the measurement are definite and distributed according to the Born rule, which is here derived and not assumed.

\section{Continuous Spontaneous Localisation and Di\'osi-Penrose models}

The two most known and studied collapse models are the Continuous Spontaneous Localisation (CSL) model \cite{pearle1989combining,ghirardi1990markov} and the Di\'osi-Penrose (DP) model \cite{diosi1987universal,penrose1996gravity}, which can be both described in terms of Eq.~\eqref{csl} with different choices of of the correlation function $D({\bf x}-{\bf y})$.\\

The CSL model, which is described by a Gaussian correlation function
\begin{equation}
D_\text{CSL}({\bf x}-{\bf y})=\frac{\lambda}{m_0^2} \exp(-|{\bf x}-{\bf y}|^2/4\rC^2),
\end{equation}
with $m_0$ being the mass of a nucleon, is a fully phenomenological model. It is characterised by two phenomenological parameters: the collapse rate $\lambda$, which determines the collapse strength for a single nucleon, 
and the noise correlation length $\rC$, whose value determines how large must be a
superposition to be suppressed. Different theoretical values have been proposed. For Ghirardi, Rimini, and Weber (GRW) \cite{ghirardi1986unified}, one has $\lambda=10^{-16}\,$s$^{-1}$ at $\rC=10^{-7}\,$m
so that an effective collapse only for macroscopic systems is guaranteed. Alternatively, 
for Adler\cite{adler2007lower}, one has $\lambda=4 \times 10^{-8\pm2}\,$s$^{-1}$ at $\rC=10^{-7}\,$m or  $\lambda= 10^{-6\pm2}\,$s$^{-1}$ at $\rC=10^{-6}\,$m, which are proposed by requiring that a collapse takes place in the mesoscopic regime.\\

The DP model, which is instead described by a correlation function proportional to the Newtonian potential
\begin{equation}
    D_\text{DP}({\bf x}-{\bf y})=\frac{G}{\hbar}\frac{1}{|{\bf x}-{\bf y}|},
\end{equation}
where $G$ is the gravitational constant, has its roots into the possible connection of the collapse with gravity. Due to the standard problems of divergence of the Newtonian potential, a Gaussian regularisation of the correlation function is implemented with the phenomenological parameter $R_0$ playing the role of the spatial cutoff. Theoretical suggestions by Di\'osi\cite{diosi1987universal} place $R_0$ around $10^{-15}\,$m  (equal to the proton radius), while Penrose \cite{ penrose2014gravitization} suggested to effectively making it equal to the width of the wavefunction of the system. With both choices, one obtains a model being free of fitting parameters. Nevertheless, we follow the recent literature that keeps the parameter $R_0$ free, with values being eventually constrained by experiments.

\section{Collapse effects at test}

The tests of collapse models are divided into two classes. The first class is that of interferometric experiments \cite{arndt2014testing}, where a superposition is created, let freely evolve and then measured. This class of experiments is the most natural one as it aims to detect the direct effect of collapse models, which is the suppression of quantum superpositions. The second class of experiments is that of non-interferometric tests\cite{carlessoNatPhys}, and it collects all the experiments that are not interferometric. Such a class focuses on different indirect effects due to the action of the collapse noise intropduced in Eq.~\eqref{csl}. Indeed, collapse models imposes a noise to the system, which consequently will behave differently from what predicted by quantum mechanics.

\subsection{Interferometric tests}

In interferometric experiments, one prepares the system in a superposition state and
then -- after some time required to the collapse effects to build up -- measures the corresponding interference pattern \cite{arndt2014testing,torovs2018bounds}. The collapse action is
determined by the reduction of the interference contrast.
For example, the reduction imposed by the CSL model to the interference pattern of a free single particle of mass $m$ over a time $t$ is given by \cite{torovs2017colored,torovs2018bounds}
\begin{equation}
    D_\text{CSL}(x)=\exp\left[
  -\lambda \frac{m^2}{m_0^2}t\left(
1-\frac{\sqrt{\pi}}{2} \frac{\erf(\tfrac{x}{2\rC})}{\tfrac{x}{2\rC}} 
  \right)  
    \right].
\end{equation}

\begin{figure} [ht]
   \begin{center}
   \includegraphics[width=10cm]{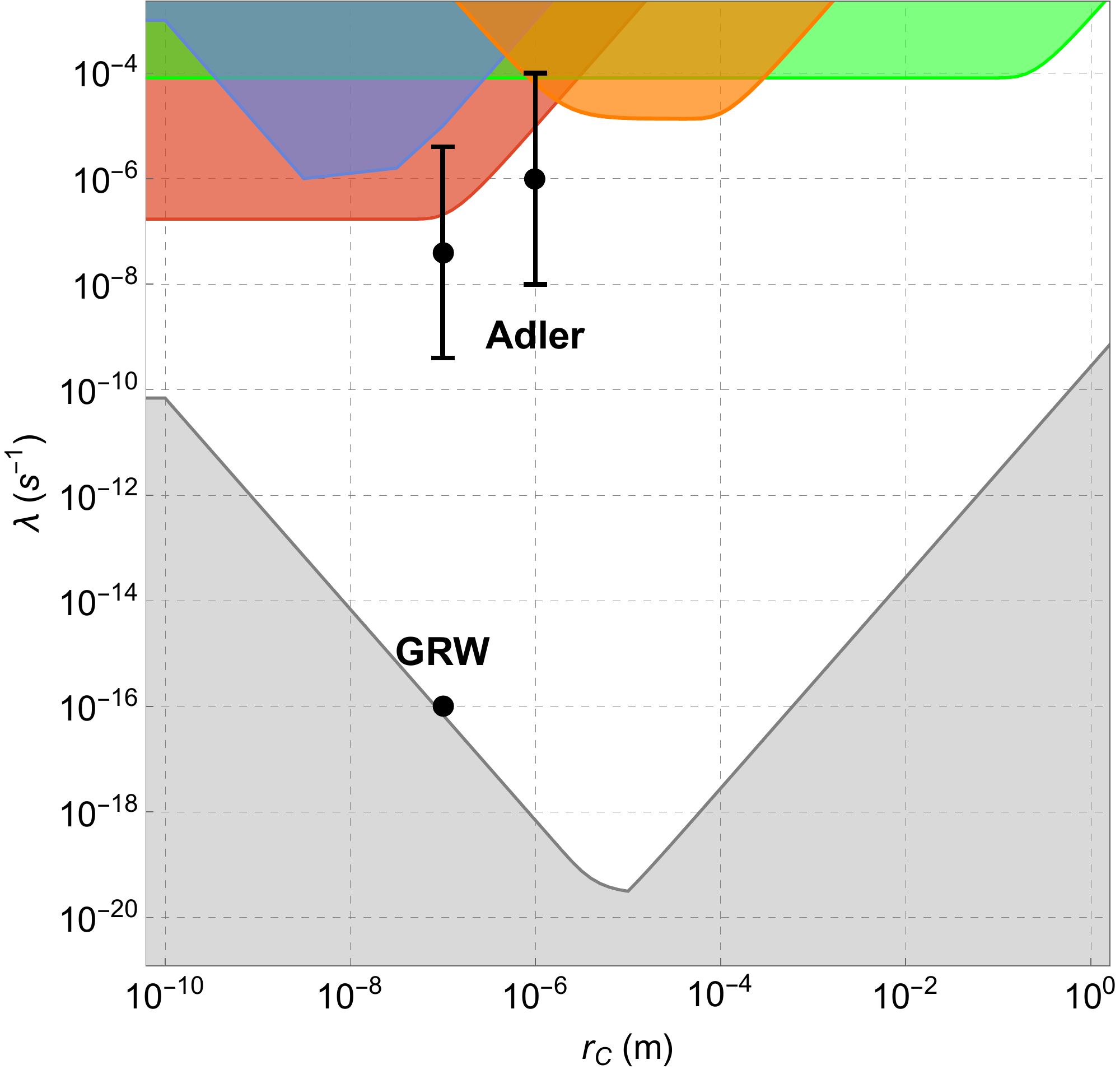}
   \end{center}
   \caption[example] 
   { \label{fig1} 
Experimental upper bounds on CSL parameters $\lambda$ and $\rC$ from interferometric experiments.  The green region is excluded from cold atoms experiment \cite{kovachy2015quantum,carlesso2019collapse}. The blue\cite{eibenberger2013matter} and red \cite{fein2019quantum} regions are excluded by molecular interferometry\cite{torovs2017colored}. The orange region is excluded from entanglement experiments with
diamonds \cite{lee2011entangling,belli2016entangling}. The theoretical values proposed by GRW\cite{ghirardi1986unified} and the ranges proposed
by Adler\cite{adler2007lower} are shown respectively as a black dot and black dots
with bars which indicate the estimated range. Finally, the light grey
area is theoretically excluded\cite{torovs2018bounds}. The white area has not been explored with interferometric experiments. Figure adapted from Ref.~\citenum{carlesso2019collapse}
}
   \end{figure}

We summarise the state of the art of interferometric experiments in Fig.~\ref{fig1} for the CSL model, where one places upper bounds on the value of $\lambda$ for a specific value of $\rC$. On the other hand, there are no substantial bounds on $R_0$ for the DP model from interferometric experiments. 

\subsection{Non-interferometric tests}

\begin{figure} [ht]
   \begin{center}
   \includegraphics[width=10cm]{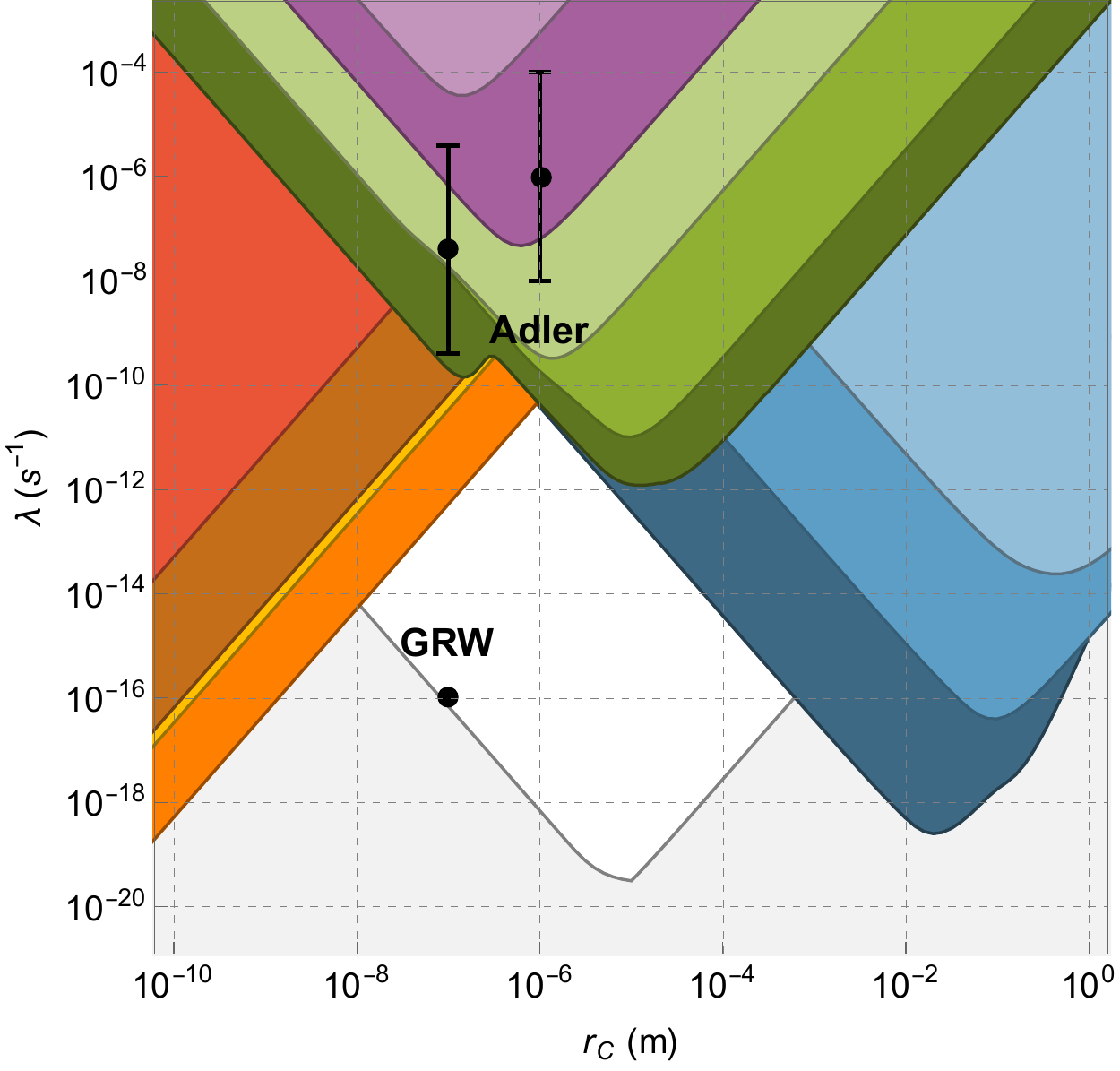}\\
   \end{center}
   \caption[example] 
   { \label{fig2} 
Experimental upper bounds on CSL parameters $\lambda$ and $\rC$ from non-interferometric experiments. 
The two purple regions are excluded by optomechanical experiments with optically levitated systems\cite{pontin2020ultranarrow,zheng2020room}. The three green regions are excluded by optomechanical experiments with cantilevers \cite{vinante2016upper,vinante2017improved,vinante2020narrowing}.
The three blue regions\cite{carlesso2016experimental,helou2017lisa,carlesso2018non} are excluded by gravitational wave detectors AURIGA \cite{vinante2006present}, LIGO \cite{PhysRevLett.116.061102} and LISA Pathfinder \cite{armano2018beyond}. The red region is excluded by a cold atoms experiment \cite{kovachy2015matter,bilardello2016bounds}. The brown region is excluded from observations of the blackbody radiation of Neptune \cite{adler2019testing}. The yellow region is excluded by phonon excitations in the CUORE experiment \cite{alduino2017projected,adler2018bulk}. The orange region is excluded by X-ray emission tests\cite{Donadi:2021tq}.
 The theoretical values proposed by GRW\cite{ghirardi1986unified} and the ranges proposed
by Adler\cite{adler2007lower} are shown respectively as a black dot and black dots
with bars which indicate the estimated range. Finally, the light grey
area is theoretically excluded\cite{torovs2018bounds}.  The white area is yet to be explored. Figure taken from Ref.~\citenum{carlessoNatPhys}.
}
   \end{figure}

\begin{figure} [ht]
   \begin{center}
   \includegraphics[width=10cm]{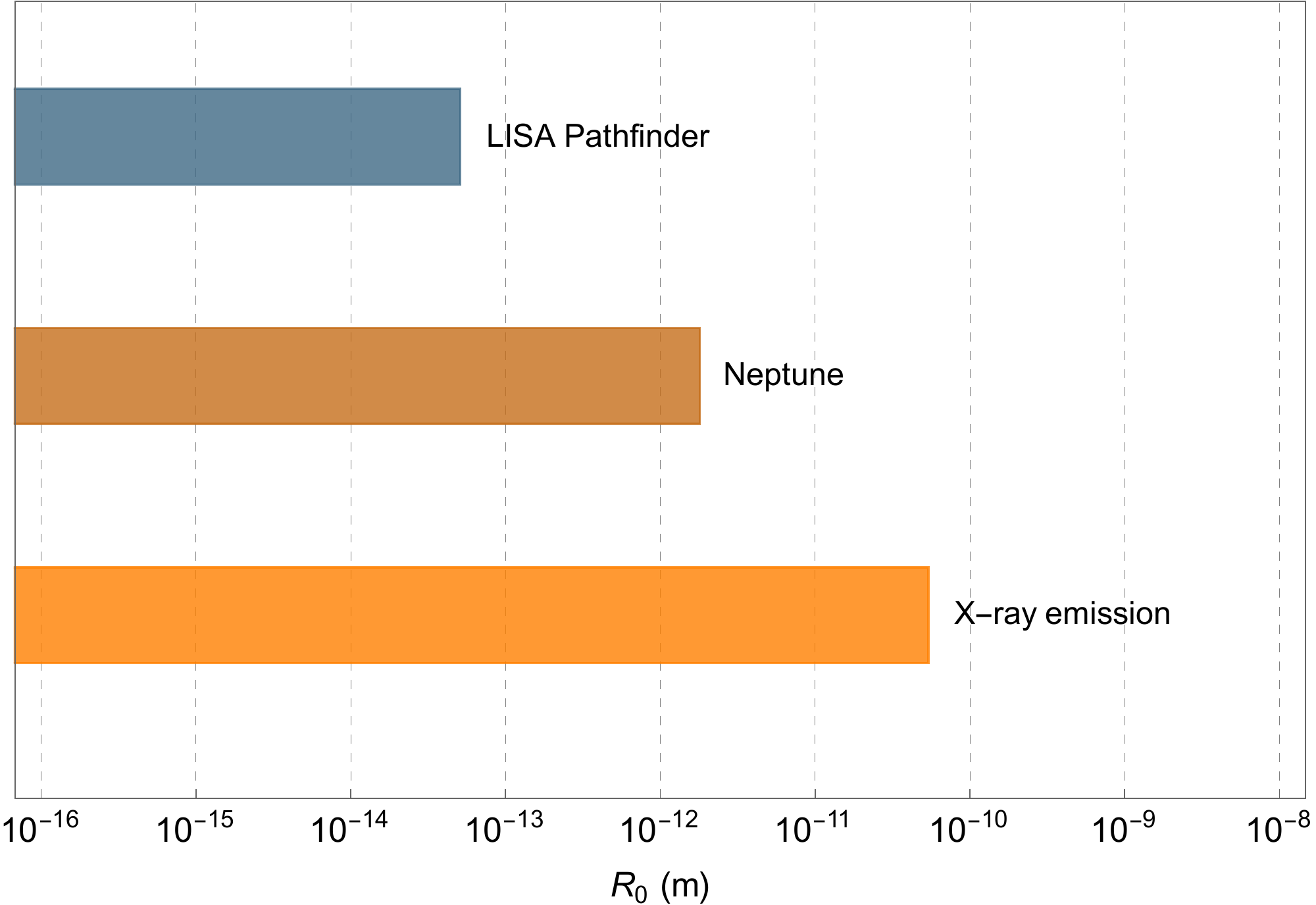}
   \end{center}
   \caption[example] 
   { \label{fig3} 
Experimental upper bounds on the DP parameter $R_0$ from non-interferometric experiments. 
The blue bound\cite{helou2017lisa} is from LISA Pathfinder\cite{armano2018beyond}.  The brown bound is  from observations  of the blackbody radiation of Neptune\cite{carlessoNatPhys}.  The orange bound is from X-ray emission tests \cite{donadi2020underground}. Figure taken from Ref.~\citenum{carlessoNatPhys}.
}
   \end{figure}

Conversely to interferometric experiments, in non-interferometric tests one can exploit different indirect effects of the action of collapse models \cite{carlessoNatPhys}. Indeed, the action of the collapse noise induces a jiggling motion to the system under scrutiny and consequently leads to an increase of its translational  (or rotational) and internal energy\cite{adler2021continuous}. The latter can be directly measured, for example in optomechanical or phonons' experiments, or it could lead to a spontaneous radiation emission if the system is electrically charged. 
For example, the heating power predicted by CSL on a system of mass $m$ is given by \cite{adler2018bulk}
\begin{equation}
    P_\text{CSL}=\frac34\frac{\hbar^2 \lambda m}{m_0^2\rC^2}.
\end{equation}
The state of the art of non-interferometric experiments\cite{carlessoNatPhys} is presented in Fig.~\ref{fig2} for the CSL and Fig.~\ref{fig3} for the DP model.\\

As one can see from the comparison of Fig.~\ref{fig1} and Fig.~\ref{fig2}, the class of non-interferometric test provides a quite stronger insight into the collapse mechanism and thus into the limits of quantum mechanics. The main reason for this is that non-interferometric experiments do not require the initial preparation of the system in a superposition state, which strongly simplifies the experimental procedure and allows the use of much more massive systems.

\section{Perspectives and challenges ahead}

New dedicated experiments are needed to further tests collapse models. They will need to achieve new levels of control of the probe mass and new levels of measurement accuracy on the collapse-induced effects.
Beside the translational degrees of freedom, that have been well exploited in several experiments, one could try to exploit also rotovibrational ones\cite{Schrinski2017,carlesso2018non}. 
To enhance the capabilities to detect collapse models effects over the hindering action of standard decoherence noises, one can think of space-based experiments\cite{kaltenbaek2016macroscopic,gasbarri2021testing,belenchia2021test,belenchia2022quantum} or in experiments providing long free-fall times as in the case of drop-towers~\cite{gierse2017fast,lotz2020tests}. Here, the probe mass can freely levitate without any external potential that will inevitably introduce
extra noises in the system’s dynamics. Although applications to cosmology have been also considered\cite{PhysRevLett.118.021102,perez2006quantum,PhysRevD.85.123001,PhysRevD.88.085020,martin2020cosmic,gundhi2021impact}, it is not yet clear how collapse models should be accounted in  relativistic situations or when gravity plays an important role~\cite{Bengochea:2020vr}. This is still an open problem.
\\

On the theoretical perspective, there are improvements that can be implemented in the modelling of the collapse dynamics. For example, both the CSL and the DP models are based on the assumption of the use of a white noise which breaks the energy conservation. Such an assumption provides a problem that is twofold: a white noise is only an approximation of physical noises, and the divergence of the energy is problematic, also for a phenomenological model. For these reasons, colored\cite{adler2007collapse,PhysRevA.90.062105} and dissipative\cite{smirne2015dissipative,PhysRevA.90.062105} 
 extensions of the CSL and DP models have been developed, although alternatives in how to model such extensions are possible \cite{dibartolomeo2023}. New phenomenological parameters $\Omega_0$ and $T_0$, being respectively a frequency cut-off describing the colored collpse noise spectrum and the temperature at which the system will eventually thermalise, are introduced. Although some experiments already provide bounds on such extensions\cite{nobakht2018unitary,carlesso2018colored,pontin2020ultranarrow,vinante2020testing}, the parameter space increases considerably (from 2D to 4D for the CSL model, and from 1D to 3D for the DP model), and requires a stronger experimental effort.\\

The technological development has recently placed the tests of collapse models within the reach of state of the art experiments. This has led
to a growing interest
of the scientific community in the field. Nevertheless, the path ahead in testing collapse models, and thus the ultimate limits of quantum mechanics, is not straightforward and requires a collective effort of the entire scientific community.

\acknowledgments 
 
MC is supported by UK
EPSRC (Grant No.~EP/T028106/1). 

\bibliography{report} 
\bibliographystyle{spiebib} 

\end{document}